\documentclass[aps,prb,preprint,superscriptaddress,showpacs]{revtex4-1}

\usepackage{graphicx,psfrag}
\usepackage[latin1]{inputenc}
\usepackage{natbib}
\usepackage{amsmath, amsthm, amssymb}
\usepackage{color}
\usepackage{comment}
\usepackage{amsmath}
\usepackage{setspace}
\usepackage{gensymb}
\usepackage{listings} 

\newcommand{\notes}[1]{}

\newcommand{\beq}{\begin{equation}}
\newcommand{\eeq}{\end{equation}}
\newcommand{\beqnn}{\begin{equation*}}
\newcommand{\eeqnn}{\end{equation*}}
\newcommand{\beqas}{\begin{eqnarray*}}
\newcommand{\eeqas}{\end{eqnarray*}}
\newcommand{\beqa}{\begin{eqnarray}}
\newcommand{\eeqa}{\end{eqnarray}}

\begin{document}


\title{Simultaneous imaging of strain waves and induced magnetization dynamics at the nanometer scale}

\date{\today}
\author{Michael Foerster*}
\affiliation{ALBA Synchrotron Light Source, 08290 Cerdanyola del Valles, Spain}
\author{ Ferran Macià*}
\email{fmacia@icmab.es}  
\affiliation{Institut de Ciència de Materials de Barcelona (ICMAB-CSIC), Campus UAB, 08193 Bellaterra, Spain}
\affiliation{Dept. of Condensed Matter Physics, University of Barcelona, 08028 Barcelona, Spain}
\author{Nahuel Statuto}
\affiliation{Dept. of Condensed Matter Physics, University of Barcelona, 08028 Barcelona, Spain}
\author{Simone Finizio}
\affiliation{Institut f\"ur Physik, Johannes Gutenberg Universit\"at Mainz, 55099 Mainz, Germany}
\affiliation{Swiss Light Source, Paul Scherrer Institut, CH-5232 Villigen PSI, Switzerland}
\author{Alberto Hernández-Mínguez}
\affiliation{Paul-Drude-Institut fur Festk\"orperelektronik, Hausvogteiplatz 5-7, 10117 Berlin, Germany}
\author{Sergi Lendínez}
\affiliation{Dept. of Condensed Matter Physics, University of Barcelona, 08028 Barcelona, Spain}
\author{Paulo V. Santos}
\affiliation{Paul-Drude-Institut fur Festk\"orperelektronik, Hausvogteiplatz 5-7, 10117 Berlin, Germany}
\author{Josep Fontcuberta}
\affiliation{Institut de Ciència de Materials de Barcelona (ICMAB-CSIC), Campus UAB, 08193 Bellaterra, Spain}
\author{Joan Manel Hernàndez}
\affiliation{Dept. of Condensed Matter Physics, University of Barcelona, 08028 Barcelona, Spain}
\author{Mathias Kl\"aui}
\affiliation{Institut f\"ur Physik, Johannes Gutenberg Universit\"at Mainz, 55099 Mainz, Germany}
\author{Lucia Aballe}
\affiliation{ALBA Synchrotron Light Source, 08290 Cerdanyola del Valles, Spain}





\begin{abstract}
	
Changes in strain can be used to modify electronic and magnetic properties in crystal structures \cite{Zeches2009,Si2016}, to manipulate nanoparticles and cells \cite{Dinga2012}, or to control chemical reactions \cite{Strasser2010}. The magneto-elastic (ME) effect---the change of magnetic properties caused by the elastic deformation (strain) of a magnetic material---has been proposed as an alternative approach to magnetic fields for the low power control of magnetization states of nanoelements since it avoids charge currents, which entail ohmic losses. Multiferroic heterostructures \cite{Zheng2004} and nanocomposites have exploited this effect in search of electric control of magnetic states, mostly in the static regime. Quantitative studies combining strain and magnetization dynamics are needed for practical applications and so far, a high resolution technique for this has been lacking. Here, we have studied the effect of the dynamic strain accompanying a surface acoustic wave on magnetic nanostructures. We have simultaneously imaged the temporal evolution of both strain waves and magnetization dynamics of nanostructures at the picosecond timescale. The newly developed experimental technique, based on X-ray microscopy, is versatile and provides a pathway to the study of strain-induced effects at the nanoscale. Our results provide fundamental insight in the coupling between strain and magnetization in nanostructures at the picosecond timescale, having implications in the design of strain-controlled magnetostrictive nano-devices.

\end{abstract}
\maketitle

Magnetization states in magnetic materials are fundamental building blocks for constructing memory, computing and further communication devices at the nanoscale. Static states such as magnetic domains are being used in non-volatile memories \cite{Akerman2005}, whereas dynamic excitations---spin-waves---might serve to transmit signals and encode information in future electronic devices \cite{Locatelli2014}. Collective magnetization states, which result from electron exchange coupling, are traditionally modified through magnetic fields created with electrical currents, giving rise to heat dissipation and stray fields. The spin-transfer-torque effect \cite{Slonczewski1996,Berger1996,Ralph2008}, which can be originated from pure spin currents offer promising pathways towards the control of magnetic states at the nanoscale without using magnetic fields. Another promising strategy for handling high-speed magnetic moment variation at the nanoscale together with low-power dissipation is the use of electric fields. Although direct effects of electric fields on magnetic states are weak, electric fields can be used to induce strain and elastic deformations in a nanoscale magnetic material that might result in changes of magnetic properties as shown mostly by static experiments \cite{Lei2013,Buzzi2013,Halley2014,Li2014,Finizio2014}.

Surface acoustic waves (SAWs) are propagating strain waves that can be generated through oscillating electric fields at the surface of piezoelectric materials. SAWs have been used to induce magnetization oscillations in magnetic materials and to achieve assisted reversal of the magnetic moment \cite{Hernandez2006, Davis2010,Scherbakov2010,Weiler2011,Weiler2012,Thevenard2016}. However, SAW induced magnetization dynamics is mostly treated as an effective variation in the magnetic energy, providing thus little information regarding the physical coupling between phononic and magnetization modes. 

In this letter we report an experimental study that probes the dynamic coupling of SAWs with the magnetization dynamics of nano-elements. The study provides a simultaneous direct observation of both strain waves and magnetization modes with high spatial and temporal resolution. Our technique combines time and spatially resolved X-ray magnetic circular dichroism (XMCD) \cite{Stohr2007} and Photoemission Electron Microscopy (PEEM)\cite{Aballe2015}. While XMCD probes the magnetization, the low-energy electrons detected by PEEM yield information on the piezo-electric potential caused by the strain wave in the piezoelectric substrate, thus providing a local measurement of strain strength. Stroboscopic XMCD and PEEM images synchronized with the SAWs allow us to correlate the local changes in magnetization with the spatial variation of the strain field.



\begin{figure}[htb!]
\includegraphics[width=0.8\columnwidth]{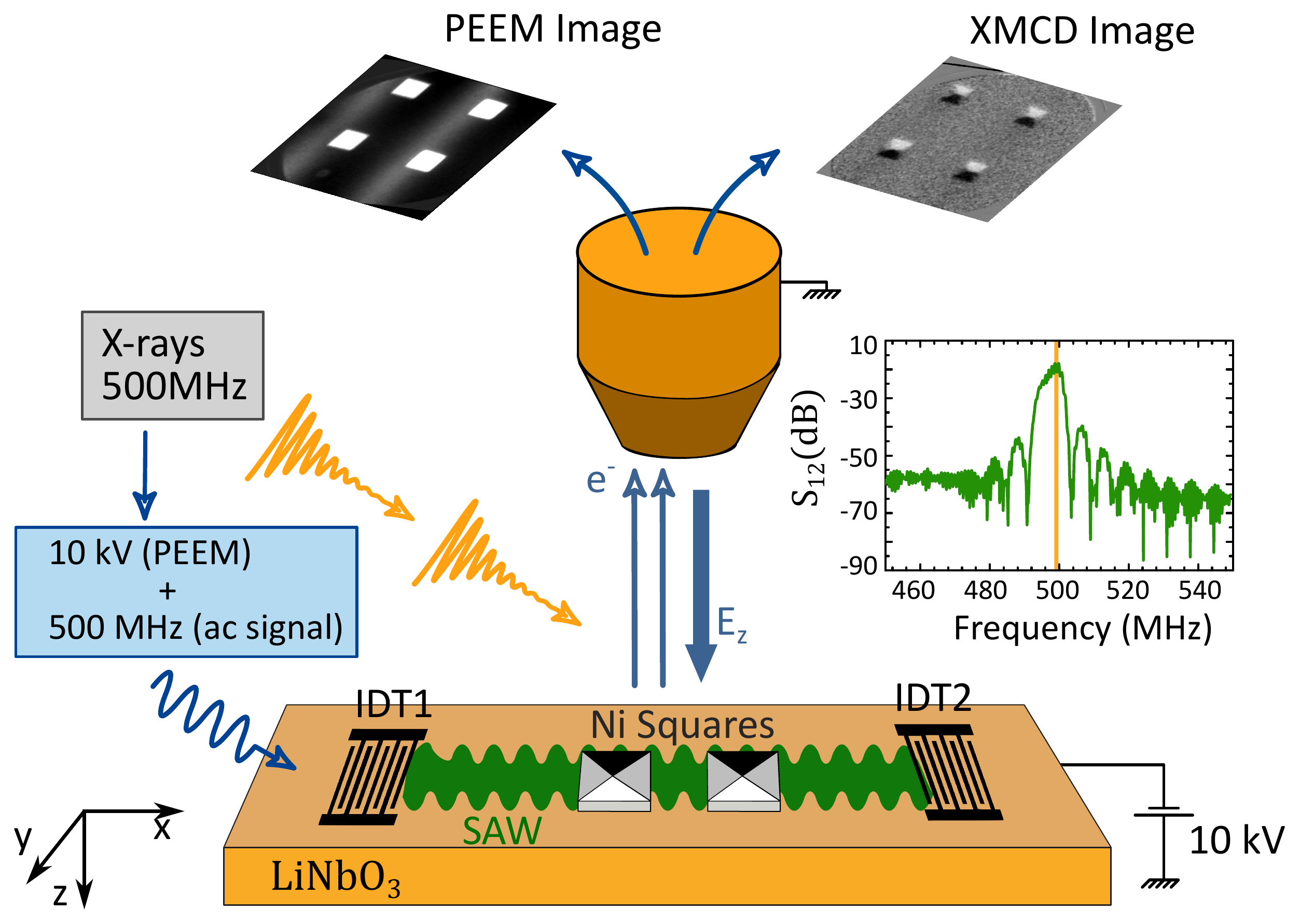}
\caption{\small{Schematic plot of the experimental set-up. Circularly polarized X-rays illuminate the sample in the form of 20 ps pulses with a repetition rate of $f_0 \approx 500$ MHz. The IDT 1 of the hybrid device receives an AC electric signal of the same frequency, which is phase locked to the synchrotron repetition rate, generating a piezoelectric surface acoustic wave (SAW) that propagates through the LiNbO$_3$ substrate and interacts with the magnetic nanostructures. The phase-resolved variation of the piezoelectric voltage at the surface sample is probed with the PEEM, as well as the magnetization contrast along the X-ray propagation direction arising from the XMCD effect. Top-left: PEEM image with a field of view of 20 $\times$ 20 $\mu$m$^2$ shows four 2$\times$2 $\mu$m$^2$ Ni squares in presence of a piezoelectric wave---black and white stripes indicate the sign of the piezoelectric voltage. Top-right: XMCD image of the same structures showing magnetic domain structures within the Ni squares. Inset: Experimental rf-power spectrum of the transmission coefficient between IDT1 and IDT2, $S_{21}$, tuned to have a maximum at $f_0$.}}
\label{fig1}
\end{figure}

A schematic plot of the measurement is shown in Fig.\ \ref{fig1}. Micrometric Nickel (Ni) squares were deposited onto piezoelectric LiNbO$_3$ substrates containing interdigital transducers (IDTs) for the excitation of SAWs. The IDTs were designed to launch SAWs of a frequency $f_{\text{SAW}}=499.654$ MHz at room temperature, which is exactly the repetition rate of X-ray bunches at the  \emph{ALBA  Synchrotron} in multibunch mode. By using an electronic phase locked loop (PLL) between the synchrotron master clock and the rf-excitation signal applied to the IDT, we achieved phase synchronization between the SAW and the X-ray light pulses illuminating the sample \cite{Foerster2016} (see, Methods and Supplementary Materials). For each phase delay between the SAW and the X-ray pulses, we recorded PEEM images that provided magnetic contrast of the sample surface through the XMCD effect. These stroboscopic measurements allowed us to reconstruct the strain wave propagation and its effect on the magnetic structures with a time resolution of $\approx 80$ ps.

\begin{figure}[htb!]
\includegraphics[width=0.8\columnwidth]{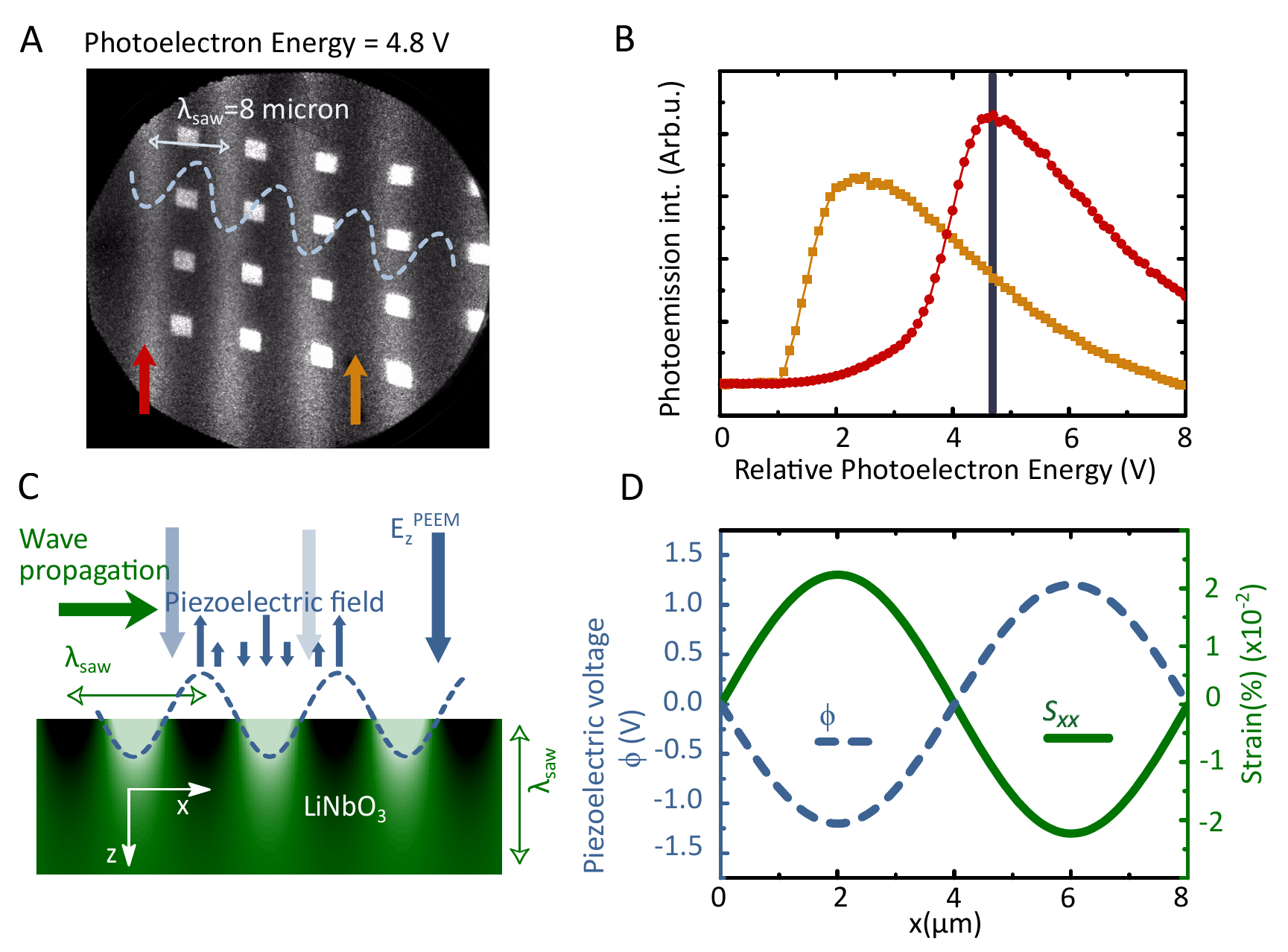}
\caption{\small{(A) PEEM image of multiple Ni squares of 2$\times$2 $\mu$m$^2$ with a field of view of 50 $\mu$m recorded at a photoelectron energy of 4.8 V. The piezoelectric voltage produces periodic dark and bright zones in the PEEM image. (B) Two photoelectron energy scans of the photoemission intensity corresponding to a bright and a dark zone (indicated in A). The vertical line indicates the electron energy at which A was acquired. (C) Schematic of the in-plane strain induced modulation in the LiNbO$_3$ caused by the acoustic wave. Blue arrows indicate the oscillating piezoelectric voltage associated with the strain modulation. (D) Calculation of the in-plane strain component (right-hand-side axis) and the piezoelectric voltage (left-hand-side axis) at the sample surface for a SAW with a 8 $\mu$m wavelength ($\lambda_{\text{SAW}}$). Calculations of the strain are done to match the measured piezoelectric voltage of Fig.\ \ref{fig2}A. The magnetization changes are driven basically by the in-plane strain component along the SAW propagation direction. The piezoelectric voltage is in phase with the in-plane strain component, albeit with an opposite sign.}}
\label{fig2}
\end{figure}

In Fig.\ \ref{fig2}A we show a PEEM image with a field of view of 50 $\mu$m containing Ni squares of 2$\times$2 $\mu$m$^2$ in presence of SAWs. We observe bright and dark stripe lines with the periodicity of the SAW excitation (wavelength, $\lambda_{\text{SAW}}\approx 8$ $\mu$m). The SAW produces a contrast in the PEEM images because the piezoelectric voltage associated with the wave shifts the energy of the secondary electrons that leave the sample surface. Imaging with a fixed phase delay and a slightly detuned (sub Hz) SAW frequency confirmed the SAW propagation direction by direct observation of the displacement of the stripes in consecutive PEEM images (see, videos in Supplementary Material). Figure\ \ref{fig2}B shows the number of secondary electrons (photoemission intensity) as a function of the electron kinetic energy, recorded by our detector at two surface areas corresponding to opposite phases of the wave, cf. the arrows in Fig.\ \ref{fig2}A. The energy shift between the two spectra corresponds to the peak-to-peak amplitude of the SAW-induced piezoelectric potential added to the 10 keV applied at the sample surface for PEEM detection (2.6 V for the rf-power used in Fig.\ \ref{fig2}A). A schematic plot of the piezoelectric SAW is presented in Fig.\ \ref{fig2}C showing the intensity of the strain modulation in the $x$-$z$ plane of a SAW propagating along $x$. Strain arises from the spatial variation of the displacements; 
the SAW decays exponentially with depth with a decay length of the order of the SAW wavelength \cite{SAW1,SAW2}. Figure \ref{fig2}C shows, as well, the electric field (blue arrows) corresponding to the modulation along $x$ of the piezoelectric voltage at the sample surface (dashed blue line).

The measurement of the amplitude of the surface electric potential associated with the SAW allows for a quantification of the strain applied to the Ni nanostructures (see, Methods). We plot in Fig.\ \ref{fig2}D the oscillation along $x$ of the piezoelectric potential with a peak-to-peak amplitude of 2.6 V (blue dashed curve) measured in Fig.\ \ref{fig2}A, together with its corresponding calculated longitudinal in-plane strain component, $S_{xx}$ (solid curve), which is the strain component responsible for the variations of the in-plane magnetic anisotropy in our structures. We notice that the piezoelectric potential, and therefore the out-of-plane electric field, $E_z = -\partial_z \phi_{\text{SAW}}$, is in phase with $S_{xx}$. Once we have shown that PEEM images provide a direct visualization of the surface potential associated to the SAW and thus a quantification of the dynamic strain, we now focus on the response originated by the SAW on the magnetic structures.

\begin{figure}[htb!]
\includegraphics[width=0.7\columnwidth]{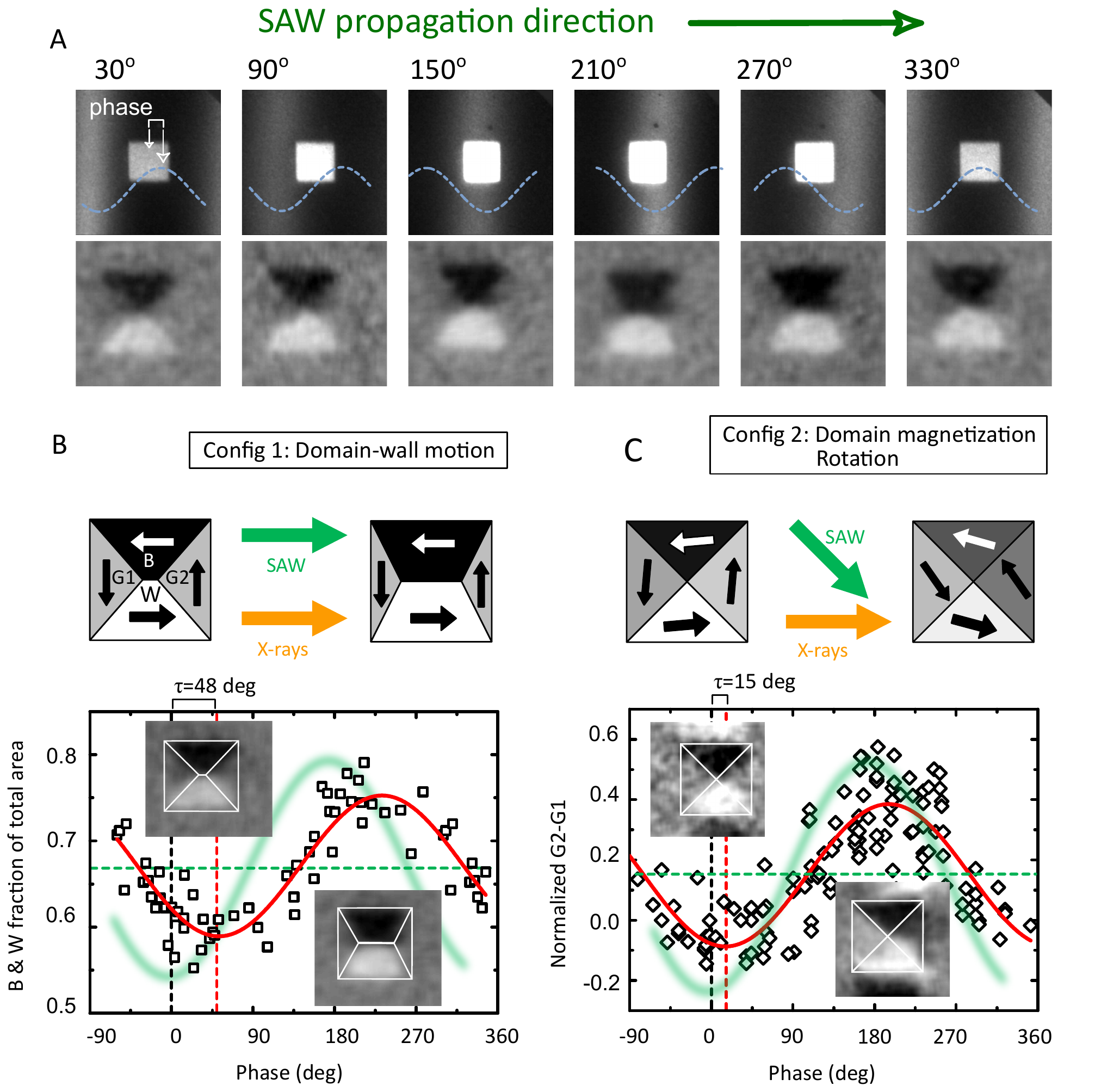}
\caption{\small{(A) PEEM (top row) and XMCD-PEEM (lower row) images of a 2$\times$2 $\mu$m$^2$ Ni square at different phases of the SAW. Images correspond to phase lapses of 60 deg (that correspond to 333 ps). PEEM images are 8$\times$8 $\mu$m$^2$; XMCD-PEEM images are 4$\times$4 $\mu$m$^2$. (B) Analysis of the domain configuration from multiple (4) Ni squares of 2$\times$2 $\mu$m$^2$ as a function of the individual phase with respect to the SAW for a configuration with square sides aligned with SAW propagation direction. Schematic plots of the effect of SAW on Ni squares shown in top panels. We computed the area of black and white domains (amount of magnetization along the X-ray incidence direction). (C) Analysis of the domain configuration from multiple (9) Ni squares of 2$\times$2 $\mu$m$^2$ as a function of the individual phase with respect to the SAW for a configuration with square sides rotated 45 degrees with SAW propagation direction. Schematic plots of the effect of SAW on Ni squares shown in top panels. We computed the intensity of the relative normalized grey domains, (I(G1)-I(G2))/(I(W)-I(B)). A best fit to the data with a sinusoidal function is plotted in red in both (B) and (C). A schematic strain wave is plotted in green (with no scale) to indicate the phase values corresponding to maximum and minimum strain corresponding to a response without delay.}}
\label{fig3}
\end{figure}

The intensity of the XMCD images is proportional to the component of the Ni magnetization along the X-ray incidence direction, represented in intensity gray scale. In order to quantify the ME-induced anisotropies, we chose polycrystalline Ni squares of $2\times2$ $\mu$m$^2$ size and 20 nm thickness, having a four-domain Landau flux-closure state \cite{Finizio2014} (see, images in Fig.\ \ref{fig3}). We first studied samples with the Ni squares' sides aligned with the SAW propagation direction. Figure\ \ref{fig3}A shows a temporal reconstruction of the effect of a SAW on a single Ni square; we plotted the direct PEEM images (top row panel) to see the SAW propagation and the PEEM/XMCD images (lower row panel) to show the magnetic domain configuration.  The dynamical process of the magnetization is precisely observed in this figure where images correspond to intervals of 333 ps (1/6 of the SAW period): gray domains are first favored (magnetization perpendicular to SAW propagation) whereas black and white domains grow at larger values of the phase (magnetization aligned with the SAW propagation).



We have analyzed the response to SAW of domain configurations from multiple squares within the same piezoelectric substrate by acquiring 20 $\mu$m$^2$ size XMCD images at different phase delays between SAW and X-ray pulses. At each phase delay, we calculated the area occupied by black and white domains in each square. This corresponds to the total area with magnetization oriented along the $x$ direction, which is the one modulated by the SAWs. Figure\ \ref{fig3}B shows a summary of the obtained values as a function of the SAW phase acting on each square (measured from the PEEM image). The ensemble of points obtained from all analyzed squares is well fitted by a sinusoidal function (red curve) with the same periodicity as the one of the SAW (green shadowed curve). We can translate the variations in the magnetic-domain configuration observed in the XMCD images into variations of magnetic anisotropy by means of micromagnetic simulations (see, Supplementary Material). In the case shown in Fig.\ \ref{fig3}B, the oscillation of the domain areas is well reproduced by a strain-induced modulation of the magnetic anisotropy of amplitude $k_{\text{ME,ac}}\approx 1$ kJ/m$^3$ superimposed to a preexisting uniaxial anisotropy of $k_{U}\approx 1.2$ kJ/m$^3$ caused by the deposition process. The static value of the preexisting uniaxial anisotropy has been confirmed by direct ferromagnetic resonance spectroscopy (FMR) measurements on the same films. We also estimated the in-plane strain at the surface of the substrate from the SAW piezoelectric potential as described in Fig.\ \ref{fig2}. For the experimental values reported in Fig.\ \ref{fig3}B, the corresponding value of the strain modulation is $S_{xx}=4.5\times 10^{-4}$.


From the correlation between in-space variations of strain and variation of magnetic anisotropy, we obtained a value for the parameter  $\beta=k_{\text{ME,ac}}/S_{xx}=2.2\times 10^6$ J/m$^3$ at 500 MHz. The value of this ME coupling coefficient is similar to the reported values measured with static strain\cite{Finizio2014}.  We expected the value $\beta$ to be similar to the static case because strain-induced changes of magnetic anisotropy are related to the modifications of electron orbitals and thus these electronic properties must respond much faster than the 500 MHz strain oscillation used in our experiment. However, we observe in Fig.\ \ref{fig3}B (and also in \ref{fig3}A) a considerable delay between the magnetization oscillation and the strain wave that amounts to $\approx$ 270 ps (phase delay of $\approx$ 50 deg), which cannot be attributed to a delay in the sound propagation from the LiNbO$_3$ substrate to the Ni structures. Phase delays can be expected if induced excitations (SAW) are coupled to internal resonances of the system \cite{Raabe2005}. We have identified three different magnetic resonance processes in the studied sample configuration with micromagnetic modeling (see, Supplementary Material), which correspond to \textit{i)} precessional
motion within the magnetic domain, \textit{ii)} domain-wall precession, and \textit{iii)} vortex motion. We found that domain-wall resonances have excitation frequencies similar to the SAW excitation frequency and thus there might be a coupling  between them that causes the delay.

Finally we explore a judiciously selected different geometry consisting of Ni squares rotated 45 degrees with respect to the SAW propagation direction in order to avoid the effect of domain wall and vortex resonances. Such configuration has four magnetic domains energetically equivalent with respect to the uniaxial varying anisotropy and thus no domain growth (shrinking) or domain-wall displacement occur; instead there is a coherent rotation of the magnetization within each of the domains. A study of this second configuration using Ni squares of the same size and thickness is plotted in Fig.\ \ref{fig3}C. Indeed, in this configuration the magnetic response is much faster, showing a sizable decrease in the delay between the SAW and the magnetization oscillation, from 270 ps down to about 90 ps (phase delay of $\approx$ 15 deg).

In summary, we have resolved simultaneously at the nanometer scale the strain caused by a SAW of $\approx 500$ MHz and the response of magnetic domains by using XMCD-PEEM microscopy, unveiling the dynamic response of the magneto-elastic effect. We found that manipulation of magnetization states in ferromagnetic structures with SAW is possible at the picosecond scale with efficiencies as high as for the static case. The magnetization dynamics is governed by the intrinsic configuration of the magnetic domains and by their orientation with respect to the SAW-induced strain, which has to be considered in the design of the magnetic devices. The described experiments offer a novel approach for the study of physical effects that depend on dynamic strain, as the concept may be applied to a wider range of research fields such as crystallography, nanoparticle manipulation, or chemical reactions.

\newpage

%

\newpage
\section*{Figures}

\newpage

\newpage

\newpage
\small

\section*{Methods}

\textbf{Sample fabrication and characterization.}
Unidirectional interdigital transducers (IDT) were patterned with photolithography and deposited with electron beam evaporation (10 nm Ti$|$ 40 nm Al$|$ 10 nm Ti) on the 128 YX cut of LiNbO$_3$ piezoelectric substrates. The transducers generate multiple harmonics of a fundamental frequency, $f_0$, which was tuned such that the 4th harmonic matches the synchrotron frequency of 499.654 MHz. 20 nm thick Ni nanostructures were defined with electron beam lithography and deposited by means of e-beam evaporation onto the piezoelectric substrate and in the acoustic path between the two IDTs (see Fig.\ \ref{fig1}). The acoustic wave transmission from IDT1 to IDT2 was characterized both at atmospheric pressure with pico-probes and in the UHV chamber in the beamline with a network analyzer. The Ni films were characterized with SQUID magnetometry to determine the saturation magnetization, $M_s=490\times 10^3$ A m$^{-1}$ and with FMR spectroscopy to determine the damping parameter and the uniaxial anisotropy,  $\alpha=0.03$ and $k_U=1200$ Jm$^{-3}$.
\\

\textbf{Experimental realization}
The experiments were performed at the CIRCE beamline of the ALBA Synchrotron Light Source \cite{Aballe2015}. The beamline employs an Elmitec spectroscopic low-energy electron and photoemission electron microscope (SpeLEEM/PEEM) operating in ultra-high vacuum. In order to generate XMCD-PEEM images, two PEEM images at the energy of the Ni $L_3$ absorption edge are acquired with opposite photon helicity (circular polarization), and then subtracted pixel-by-pixel to provide images with XMCD magnetic contrast as intensity. Samples were mounted on in-house designed sample holders \cite{Foerster2016} and the IDT contacted with wire-bonds. IDT were several mm away from the sample center and the Ni nanostructures, thus allowing to screen them from the high electric field of the objective lens by the raised sample holder cap. After introduction in vacuum each sample was degassed at low temperature ($<$ 100 C) for at least 1 h in order to reduce the risk of arcs between sample (at high voltage) and microscope objective (at ground). A reduced acceleration voltage of 10 kV (standard is 20 kV) was used to further reduce the risk of discharges. The beamline intensity was adjusted in order to avoid excessive surface charging of the LiNbO$_3$ substrate. For the synchronized excitation, the digital timing signal provided by the ALBA timing system \cite{Matilla2012} was converted into a phase locked 499.654 MHz (referred to as 500 MHz throughout the text) analog signal with a Keysight EXG Vector signal generator (model N5172B with option 1ER). The phase with respect to the master clock and the amplitude of the signal can be adjusted at this level as the experiment requires. The analog signal is then transmitted by a custom optical fiber system into the PEEM high voltage rack and amplified \cite{Foerster2016}. The phase or temporal resolution depends on the size of the zone analyzed as the phase changes with $\lambda$ approx 8 $\mu$m. However, an upper limit of the total temporal smearing (electronic jitter plus photon distribution from bunch length and bunch dephasing) for a small enough zone can be derived from the sharpest possible step feature of the SAW that can be resolved, which corresponds to ca. 80 ps. This is a good value for time resolved PEEM, which is helped by the continuous electrical AC excitation of a resonator structure (IDT), for which jitter is more easily controlled, but also reflects the quality of the ALBA beam in multi-bunch mode. All data presented was taken in thermal equilibrium.  When the SAW were switched on, a slow (time scale of minutes up to tens of  minutes) drift in the PEEM imaging showing small changes in the SAW wavelength from LNO surface was observed and indicated a change in temperature. Data was taken after a long period of thermalization and we compared snapshots  at different instants within the 2 ns SAW cycle.
\\

\textbf{Piezoelectric voltage to strain conversion}.
We calculated the amplitude of the piezoelectric potential and strain tensor components by numerically solving the coupled differential equations of the mechanical and electric displacement for an acoustic wave propagating along the $x$ direction of a semi-infinite 128 Y-cut LiNbO$_3$ substrate. To obtain surface modes, we looked for solutions that decay towards $z>0$ and satisfy the stress and electric displacement boundary conditions at the surface, $z=0$. We have used the same SAW wavelength as the experiment, and have selected power density so that the amplitude of the simulated piezoelectric potential at $z=0$ coincides with the measured one \cite{paramLiNbO}.

\textbf{Micromagnetic simulations}.
Numerical simulations were performed using a MuMax3 code \cite{mumax}on a graphics card with 2048 processing cores. A full code is appended in the Supplementary Material. We considered a two-dimensional layer and integrated the Landau-Lifshitz-Gilbert-Slonczewski equation to describe the magnetization dynamics. We computed different sample geometries with a resolution of 4 nm in the grid size. Thermal effects are neglected. The parameters of the magnetic layer, a Nickel film, were taken from the sample characterization: saturation magnetization $M_s=490\times 10^3$ A m$^{-1}$, Gilbert damping constant $\alpha=0.03$, and exchange constant $10^{-11}$ Jm$^{-1}$. (The exchange constant does not affect significantly the simulations. The damping constant plays a role for values larger than 0.1 inducing a delay of more than 5 deg. under a SAW excitation of 500 MHz. Lower damping values do not affect neither the efficiency nor the delay times but sharpen the internal resonances). A fixed uniaxial anisotropy is introduced in the simulations with a value $k_U=1200$ J m$^{-3}$ with an additional oscillating term of $k_{ME,ac}$ having the wavelength set by the SAW ($\lambda_{\text{SAW}}\approx$ 8 $\mu$m); simulations covered frequencies from tens of MHz to tens of GHz.

\section*{Acknowledgements}

The authors thank Jordi Prat for technical help on the beam line and with the data analysis, Hermann Stoll and Rolf Heidemann for advice on electronics, Abel Fontsere, Bernat Molas and Oscar Matilla from Alba electronics for the development of the 500 MHz synchronous excitation setup, and Werner Seidel from PDI for assistance in the preparation of the acoustic delay lines on LiNbO3. The project was supported by the ALBA in-house research program through IH2015PEEM and the allocation of in-house beamtime as well as with proposal 2016021647. FM acknowledges financial support from the Ramón y Cajal program through RYC-2014-16515. FM and JF acknowledge support from MINECO through the Severo Ochoa Program for Centers of Excellence in R\&D (SEV-2015-0496). Funding from MINECO through MAT2015-69144 (JMH, NS and FM) and MAT2015-64110 (LA and MF) is acknowledged. 
SF and MK aknowledge Graduate School of Excellence  Materials Science in Mainz (Grant No. GSC 266), the Swiss National Science Foundation (SNF), The German Research Foundation DFG (TRR 173 Spin+X), the ERC (ERC-2014-PoC 665672 MULTIREV), The EC (NMP3-LA-2010 246102 IFOX, FP-PEOPLE-2013-ITN 608003 WALL) and the Center for innovative and Emerging Materials at the Johannes Gutenberg Universit\"at Mainz. SF acknowledges the financial support from the EU Horizon 2020 Project MAGicSky (Grant No. 665095)

\section*{Author contributions}
MF conceived the PEEM experiment with input from SF, JF and MK. MF, FM and LA planned and directed the project. LA, MF, FM, NS, SL, JMH, and SF performed XPEEM measurements. LA, MF, FM, and SF analyzed the data. NS, JMH and FM performed micro-magnetic simulations. AH and PS designed, prepared and characterized frequency tuned IDT on LiNbO$_3$, SF prepared the Ni microstructures. All authors discussed the results and contributed to drafting the manuscript.

\newpage

\huge{Supplementary Materials}
\normalsize

\section*{Videos Description}
\begin{itemize}
	\item \textit{STVscan.avi}: The SAW produce a contrast in the PEEM images because the piezoelectric voltage associated with the strain wave shifts the energy of the secondary electrons. Thus, bright and dark stripe lines with the periodicity of the SAW excitation (wavelength, $\lambda_{\text{SAW}} \approx$ 8 $\mu$m) appear in the PEEM images. We compiled a video with PEEM images corresponding to different electron kinetic energy (that were controlled in our detector). Note the contrast inversion during the scan corresponding to the ranges highlighted in Figure 2b. A single image is presented in Fig.\ 3A in the main manuscript.
	
	\item  \textit{Detuned.avi} We recorded PEEM images with a SAW frequency having a small (sub-Hz) detuning with respect to the synchrotron bunch frequency to confirm the SAW propagation direction by direct observation of the displacement of the stripes in the PEEM images. This video presents a 0.01 Hz detuning both positive and negative that confirm the wave propagation.
	
	\item \textit{XMCD.avi} PEEM and PEEM/XMCD images are combined in a video that shows the simultaneous evolution of the piezoelectric voltage and the magnetic domain configuration in the hybrid sample of LiNbO$_3$ with Ni squares.
	
\end{itemize}

\section*{Micromagnetics results}

We modeled the dynamic anisotropy variations in the Ni nanostructures with micromagnetic simulations using the open-source MuMax3 code \cite{mumax} on a graphics card with 2048 processing cores. Simulation parameters are reported in the Methods section and a simplified code is listed in the following section.

\begin{figure}[htb!]
	\includegraphics[width=0.8\columnwidth]{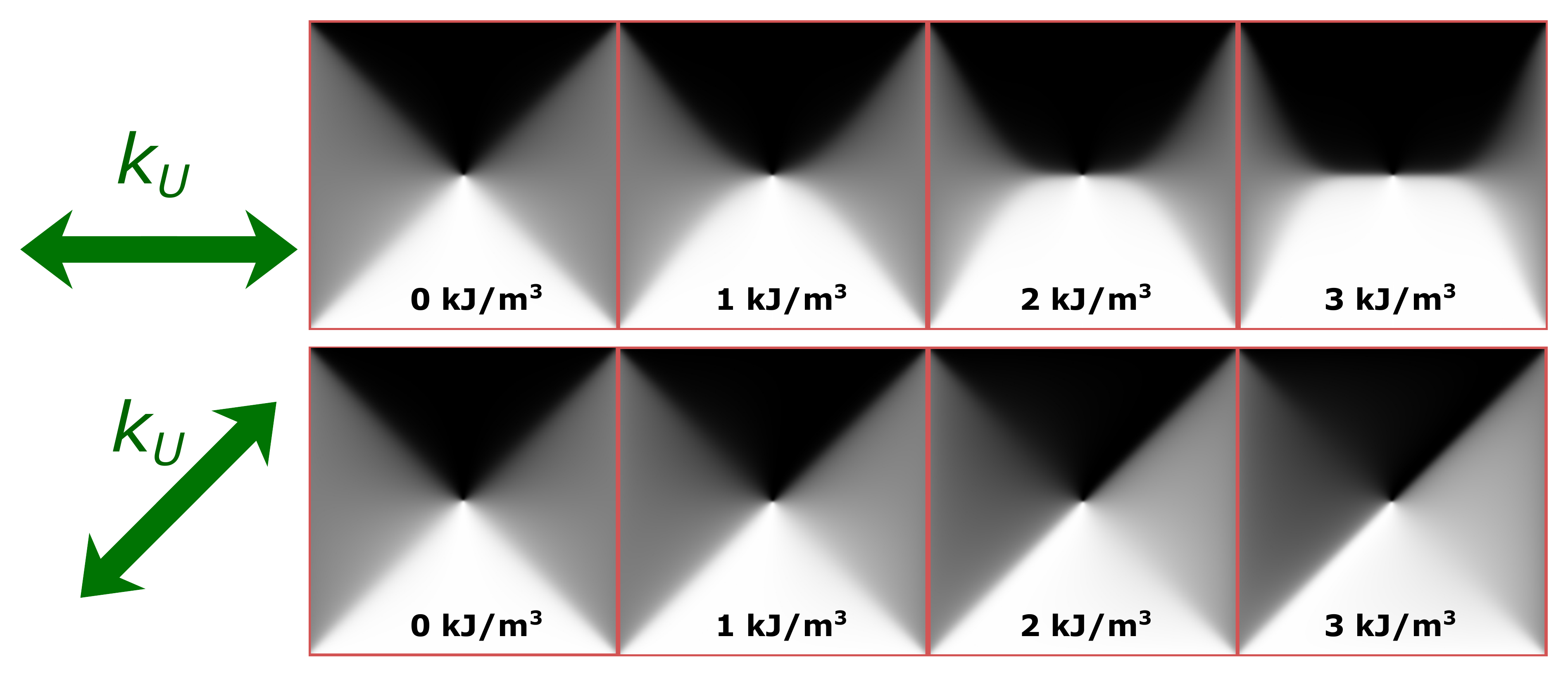}
	\caption{\small{Magnetic domain configurations corresponding to different uniaxial anisotropies. Top panels show magnetic-domain configuration on a Ni square $2\times2$ $\mu$m$^2$ with 20 nanometer in thickness, considering that the anisotropy axis is along the square sides. Lower panels present the case where anisotropy axis is along square diagonal on the same micrometric structures}}
	\label{s1}
\end{figure}

We can estimate the induced anisotropy of a given magnetic-domain configuration and thus we can translate the variations observed in the experiments into variations of the magnetic anisotropy. Magnetic domain configurations corresponding to different uniaxial anisotropies are shown in Fig.\ \ref{s1}. The top panels show magnetic-domain configuration on a Ni square $2\times2$ $\mu$m$^2$ considering that the anisotropy axis is along the square sides whereas the lower panels present the case where anisotropy axis is along the diagonal of the square.
\\

The studied magnetic nanostructures have internal resonances in the magnetic domains, in the domain walls, and even in the vortex formed in the center of the Landau flux closure states. Raabe \emph{et al.} \cite{Raabe2005} experimentally identified three different dynamic processes---with different timescales---in a Ni micrometric squares under a short magnetic pulse; i) precessional motion within the magnetic domain, ii) domain-wall precession and iii) vortex motion. In our experiment we studied two configurations that precisely can serve to separate the dynamics involving domain wall motion (square sides aligned with the SAW) from dynamics that only involves magnetization rotation within a domain (square diagonal aligned with the SAW).

\begin{figure}[htb!]
	\includegraphics[width=0.8\columnwidth]{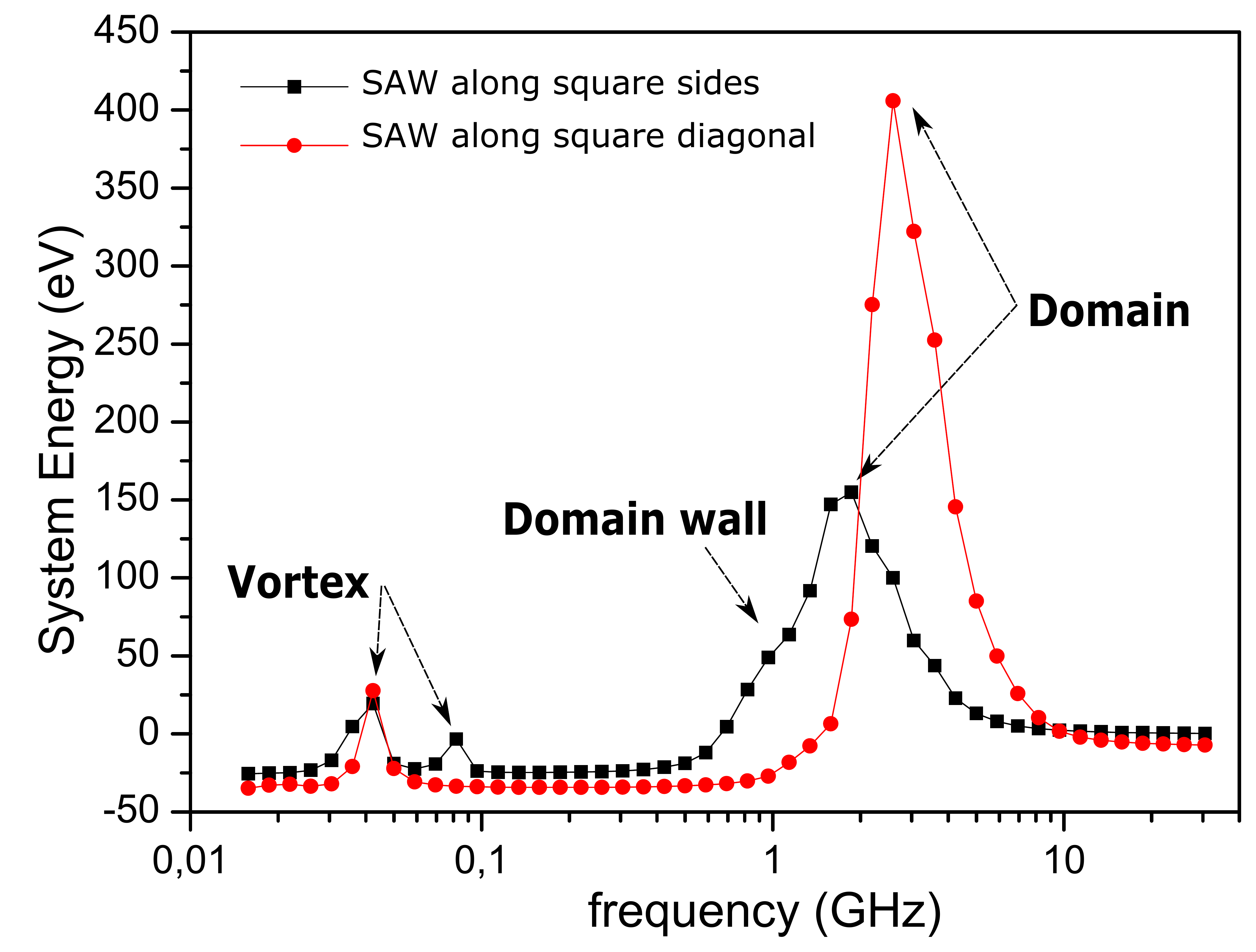}
	\caption{\small{ Energy of the magnetic domain configuration of a Ni square $2\times2$ $\mu$m$^2$ with 20 nanometer in thickness as a function of the SAW frequency perturbation. The anisotropy is modulated with a fixed wavelength $\lambda_{\text{SAW}}=8$ $\mu$m to emulate the effect of the SAW.}}
	\label{s2}
\end{figure}

We have introduced in the simulations a time varying anisotropy with a fixed wavelength $\lambda_{\text{SAW}}=8$ $\mu$m and measured the magnetic response of the Ni squares under different frequencies of the oscillating anisotropy. In order to quantify the dynamic state of the Ni squares we calculated the averaged magnetic energy at each frequency. Fig.\ \ref{s2} plots the system energy in the Ni square as a function of the SAW frequency (an oscillating uniaxial anisotropy) for the two studied configurations: SAW aligned with square sides (in black) and SAW aligned with square diagonal (in red).

Resonance frequencies produce variation in the magnetic-domain configuration having a higher energy and can be observed as peaks in the plot of Fig.\ \ref{s2}. The configuration with the anisotropy aligned along the square sides presents a richer response to SAW frequencies; two resonance peaks at $\approx 45$ MHz and $\approx 90$ MHz corresponding to vortex motion, one peak at $\approx 1$ GHz corresponding to the domain-wall resonance and one peak at $\approx 2$ GHz corresponding to domain resonance. The configuration with the anisotropy aligned with the square diagonal is much simpler because the domain walls remain static and only the magnetization within the domains varies; there is a resonance at $\approx 45$ MHz corresponding to the vortex oscillation and a resonance at $\approx 3.5$ GHz corresponding to the domain resonance.

\newpage
\newpage
\section*{Micromagnetics code}

\begin{spacing}{0.8}
	\emph{
		{\footnotesize
			\\
			// mumax3 is a GPU-accelerated micromagnetic simulation open-source software\\
			// developed at the DyNaMat group of Prof. Van Waeyenberge at Ghent University.\\
			// The mumax3 code is written and maintained by Arne Vansteenkiste.\\
			\\
			//GRID\\
			CellSize:=4.e-9\\
			NumCells:=512\\
			SetGridSize(NumCells, NumCells, 1)\\
			SetCellSize(CellSize, CellSize, 20.e-9)\\
			\\
			Setgeom(universe())\\
			\\
			//MATERIAL PARAMETERS FOR STANDARD Ni\\
			Msat=490e3\\
			Aex=1e-11\\
			Alpha = 0.03\\
			\\
			//INITIAL MAGNETIZATION STATE\\
			m = vortex(1,1)\\
			\\
			//REGIONS\\
			MaxRegion:=200\\
			CellsPerRegion:=NumCells/MaxRegion\\
			RegionWidth:=CellsPerRegion*CellSize\\
			SampleCenter:= CellSize*NumCells/2.\\
			\\
			for i:=0; i$<$=MaxRegion; i++ \{ \\
			defregion(i, xrange(i*RegionWidth-SampleCenter,1))\\
			\}\\
			\\
			//DEFINING ANISOTROPY VECTOR\\
			for i:=0; i$<$=MaxRegion; i++\{\\
			AnisU.SetRegion(i,vector(1.,0.,0.))\\
			\}\\
			\\		
			//DEFINING ANISOTROPY CONSTANT\\
			\begin{tabular}{ll}
				freq:=500000. 	& //SAW freq \\
				Kuav:=1.2e3 	&//Nominal Anisotropy\\
				Kumod:=1.e3	& \\
				Lambd:=4. 	&//in sample width units\\
			\end{tabular}
			\\
			for i:=0; i$<$=MaxRegion; i++\{\\
			ku1.SetRegion(i, Kuav + Kumod * cos( 2*pi*((i* 1/MaxRegion -0.5)/Lambd - freq*t)))\\
			\}\\
			\\
			relax()\\
			run(20e-9)
		}
	}
	
\end{spacing}

\end{document}